\documentclass[12pt]{iopart}
\usepackage{ulem}
\usepackage{color}
\usepackage{graphicx}
\usepackage{epsfig}
\newcommand{\be}{\begin{equation}}
\newcommand{\ee}{\end{equation}}
\newcommand{\ba}{\begin{eqnarray}}
\newcommand{\ea}{\end{eqnarray}}
\newcommand{\nn}{\nonumber\\}
\begin{document}
\title{Heavy-quark transport coefficients in a hot viscous quark-gluon plasma medium} 
\author{Santosh K. Das $^{1,2,3}$, Vinod Chandra $^4$, and Jan-e Alam $^5$}
\address{$^1$ Department of Physics,Yonsei University, Seoul, Korea}
\address{$^2$ Department of Physics and Astronomy, University of Catania, Via S. Sofia 64, I-
95125 Catania, Italy}
\address{$^3$ Laboratori Nazionali del Sud, INFN-LNS, Via S. Sofia 62, I-95123 Catania, Italy}
\address{$^4$ Istituto Nazionale di Fisica Nucleare (INFN) -Sezione di Firenze, Via G. Sansone 1, I-50019 Sesto F.no (Firenze), Italy}
\address{$^5$ Variable Energy Cyclotron Centre, 1/AF, Bidhan Nagar , Kolkata - 700064}
\ead{chandra@fi.infn.it}
\date{\today}
\begin{abstract}
The heavy-quark (HQ) transport coefficients have been estimated for a viscous 
quark-gluon plasma medium, utilizing a recently proposed quasi-particle description based on a realistic
QGP equation of state (EoS). Interactions entering through the equation of state significantly suppress 
the temperature dependence of the  drag coefficient of QGP  as compared to that  of an ideal
relativistic system of quarks and gluons. Inclusion of shear and bulk viscosities 
through the corrections to the thermal phase space factors of the bath particles 
alters the magnitude of the drag coefficient, and the enhancement is significant at lower temperatures. 
In the competition between the effects of the EoS and dissipative corrections through phase space 
factors; the former eventually dictate how the drag coefficient would behave as a function of temperature, 
and how much quantitatively digress from the ideal case. The observations suggest significant impact of both the realistic equation of state,
and the viscosities, on the HQs transport at RHIC and the LHC collision energies. 

\vspace{2mm}
\end{abstract}
  \vspace{2mm}
\maketitle

\section{Introduction}
Experimental heavy-ion collision (HIC) program at Relativistic Heavy Ion Collider
(RHIC) indicates the production of a liquid like state of the matter 
where the properties of the system is governed by quarks and gluons. 
Such a state of matter is called as quark-gluon plasma (QGP). Two of the most 
striking finding of the RHIC which lead to the imposition of strongly interacting liquid-like
behavior are {\it viz.}, the observed large 
elliptic flow and the phenomenon of jet quenching~\cite{expt}. 
Preliminary results from LHC~\cite{alice,alice1,hirano} 
confirm these observations. 
The strongly coupled picture of the QGP is found to be consistent  with the 
lattice simulations of the hot QCD (Quantum Chromodynamics) 
equation of state (EoS)~\cite{leos_lat,leos1_lat}, which predict 
a strongly interacting behavior even at temperatures, $T\sim 2T_c$,
where $T_c$ is the temperature for the quark-hadron transition.
Therefore, it calls for the adaption of a realistic equation of states for the QGP (such as obtained form 
lattice QCD) in the 
theoretical and phenomenological investigations regarding its bulk and transport properties.
We shall closely follow this point of view in the present analysis.

Hadrons containing HQs ($c$, $\bar{c}$, $b$, or $\bar{b}$)
are of great interest in investigating the properties of the QGP, since their momentum spectrum 
get significantly modified while traveling through QGP. This fact has been 
reflected in the particle spectra at RHIC, and LHC energies.
Further, HQ thermalization time is larger than gluons and light quarks,
and they do not constitute the bulk of the QGP. Since their formation occurs in the 
early stages of the collisions, they can travel through the thermalized QGP medium, and can 
retain the information about the interaction with them very efficiently. For instance,
it is pertinent to ask whether a single $c \bar{c}$ can stay together long enough to 
form a bound state (say $J/\psi$) at the hadronization state. To address this,
one requires to study the dynamics of the  HQs propagating through the QGP. 

One can envisage the HQ transport in the QGP medium as follows.
The non-equilibrated HQs travel in the equilibrated QGP medium, therefore,  
the problem can be studied within the framework of Langevin dynamics~\cite{fp}. This is to say that the 
HQs undertake a random motion in the heat bath of the equilibrated QGP. Recall, that 
the QGP goes through a hydrodynamic evolution process before it reaches the hadronization and 
subsequently the hadrons freeze-out. The pertinent question to ask is, whether a HQs maintain equilibrium during this 
entire process of evolution or not. It has been observed ~\cite{jane} within the ambit of Langevin dynamics 
and pQCD (perturbative QCD) that they may not achieve the equilibrium for the 
 RHIC and LHC collision conditions. 

The nuclear modification factor of the HQ, $R_{AA}$ encode the effects of the medium through
the modified momentum distribution due to their interaction with the bath particles. 
It has been 
observed that their energy loss in the QGP due to gluon radiation is  insufficient to  
describe the medium modification of the spectrum~\cite{mooreteaney,bair,jeon}. 
Therefore, one has to look at the collisions since they have different fluctuation 
spectrum than radiation, and might contribute significantly as indicated 
earlier~\cite{coll,akdm}. The collisional 
effects can be captured well in the HQ drag and diffusion coefficients which have been 
calculated within weak coupling QCD by several authors. The formalism, and details are offered 
in Sec. 2.1.

The temperature, $T$ and chemical potential, $\mu_B$ dependence of the 
drag and diffusion coefficient enter through the thermal distributions of gluons and light quarks.
In the present case, we ignore the $\mu_B$ dependence in view of the fact that the QGP produced at RHIC 
and LHC energies at the mid-rapidity region has negligibly small net baryon density. 
Therefore, one has to implement the realistic QGP EoS 
in terms of appropriate form of the thermal distribution functions. 
Lattice QCD EoS (LEoS) may be a good choice for the 
description of the QGP.  Another important aspect is the 
viscosities (both shear and bulk) of the QGP.  
One also has to include the viscous modifications in 
the distribution functions by maintaining the near equilibrium picture so that the mathematical formalism of 
HQ transport remain intact. This is possible since the QGP posses small shear  and
bulk viscosities for temperature, $T>T_c$ ($T_c$ is the quark-hadron transition).
Both these aspect have been included in the present work 
while studying the temperature dependence of HQ drag and diffusion coefficients in the QGP.
Discussion regarding the former is offered in Sec. 2.2, and the latter in 2.3. 
Our results suggest that both of these modifications have significant impact on the temperature dependence of 
HQ transport coefficients. Interestingly, the inclusion of interaction through the EoS 
modulate the transport coefficients quite significantly as compared to those obtained by employing an ideal 
EoS (assuming the QGP as a relativistic gas of non-interacting 
quarks and gluons). 

The paper is organized as follows.
In section 2.1, formalism of HQ drag and diffusion 
has been presented. Section 2.2 deals with the quasi-particle 
description of hot QCD and modeling of thermal 
distribution functions of gluons, and quarks which constitute 
hot QGP medium. Section 2.3 deals with the viscous modifications to 
the thermal distribution functions. In Section 3, results and
discussions have been presented for the drag and diffusion coefficients 
with the LEoS which may be treated a realistic EoS for the QGP . 
Section 4 is devoted to summary and conclusions.

\section{Heavy-quark drag and diffusion in the viscous QGP}
The special role of HQs to characterize QGP  stem 
from the fact that  they are produced very early in the collisions and
remain extant throughout the evolution and hence can witness the 
entire evolution of the system. Moreover, HQs do not constitute
the bulk part of the system because their masses is considerably larger than 
the temperature attainable in HIC at RHIC and LHC. 

\subsection{Formalism of HQ drag and diffusion}
In the present work, we consider the elastic interaction experienced 
by HQs while propagating through the QGP. 
For the process, $c(p) + l(q) \rightarrow c(p^\prime) + l(q^\prime)$ 
($l$ stands for  light quark, anti-quark and gluon),
the drag, $\gamma$ can be calculated by using the following 
expression~\cite{BS,all}:
\begin{equation}
\gamma=p_iA_i/p^2
\end{equation}
where $A_i$ is given by 
\begin{eqnarray}
A_i=\frac{1}{2E_p} \int \frac{d^3q}{(2\pi)^3E_q×} \int \frac{d^3p^\prime}{(2\pi)^3E_p^\prime×}
\int \frac{d^3q^\prime}{(2\pi)^3E_q^\prime×}  \nonumber \\ \frac{1}{g_Q} 
\sum  \overline{|M|^2} (2\pi)^4 \delta^4(p+q-p^\prime-q^\prime) \nonumber \\
{f}(q)(1\pm f(q^\prime))[(p-p^\prime)_i] \equiv \langle \langle
(p-p^\prime)\rangle \rangle
\label{eq0}
\end{eqnarray}
$g_Q$ being the statistical degeneracy of the HQ propagating through the medium.
The expression for $\gamma$ indicates that the drag coefficient is the
measure of the thermal average of the momentum transfer, $p-p^\prime$ due to
interaction weighted by the square of the invariant amplitude, $\overline{\mid M\mid^2}$.
The factor $f(q)$ denotes the thermal phase space for the particle
in the medium. We shall see in the subsequent subsection that $f(q)$
will involve three types of thermal phase space distribution 
corresponding to the  gluons ($g$), light-quarks ($q\equiv$ up and down) and 
the strange quarks ($s$) and their anti-quarks. 
Therefore, $f(q)$ jointly denote these three sectors as,
\begin{equation}
f(q)\equiv \lbrace f_g, f_q, f_s \rbrace, 
\end{equation}
in the absence of any dissipative effects.
It may be recalled here that for zero baryonic chemical potential 
the thermal phase space for quarks and anti-quarks are same.
In the presence of dissipation, we denote it as,

\begin{equation}
f(q)\equiv \lbrace f_{gg}, f_{qq}, f_{ss} \rbrace.
\end{equation}
We shall discuss these distributions in the subsequent subsections.

Similar to drag, the diffusion coefficient, $B_0$ can be defined as:
\begin{equation}
B_0=\frac{1}{4}\left[\langle \langle p\prime^2 \rangle \rangle -
\frac{\langle \langle (p.p\prime)^2 \rangle \rangle }{p^2}\right]
\label{diffusion}
\end{equation}

With an appropriate choice of ${\cal F}(p^\prime)$
both the drag and diffusion co-efficients can be evaluated from
the following expression:
\begin{eqnarray}
<<{\cal F}(p)>>=\frac{1}{512\pi^4×} \frac{1}{E_Q} \int_{0}^{\infty} \frac{q^2 dq d(cos\chi)}{E_q}{f}(q) \nonumber \\
\frac{w^{1/2}(s,m_Q^2,m_p^2)}{\sqrt{s}} \int_{1}^{-1} d(cos\theta_{c.m.})  \nonumber \\ 
\frac{1}{g_Q} \sum  \overline{|M|^2} \int_{0}^{2\pi} d\phi_{c.m.} {\cal F}(p^\prime)
\label{transport}
\end{eqnarray}
where\, $s$ is the Mandelstam variable, $E_p=\sqrt{p^2+m_Q^2}$ is the energy of the 
HQ, $E_q$ is the energy of the bath particle, $\chi$ is the polar angle of $\vec{q}$ 
and $w(a,b,c)=a^2+b^2+c^2-2ab-2bc-2ac$, is the triangular function. Note that the
Bose enhancement and Pauli suppression of phase space is neglected in Eq.~\ref{transport}. 

\subsection{The quasi-particle description of hot QCD}
Now we discuss the quasi-particle nature of the hot QCD medium which is employed 
recently in~\cite{chandra_quasi}.
This description has been developed in the context of the recent (2+1)-flavor lattice QCD EoS~\cite{cheng} 
at physical quark masses. There are more recent lattice results with the improved 
actions and refined lattices~\cite{leos1_lat}, for which we need to re-look the model 
with specific set of lattice data specially to define the effective gluonic degrees of freedom.
This is beyond the scope of the present analysis. Henceforth, we will stick to the set of lattice data
utilized in the model described in~\cite{chandra_quasi}.

The model initiates with an ansatz that
the LEoS can be interpreted in terms of non-interacting
quasi-partons having gluon and quark effective fugacities  $z_{g}$ and $z_q$ 
respectively which encode all the interaction effects. 
In this approach, the hot QCD 
medium is divided into two sectors, {\it viz.}, the effective gluonic sector, and 
the matter sectors (light and strange quark). The former refers to the contribution of 
gluonic action to the pressure which also involves contributions 
from the internal fermion lines. On the other hand, latter involve interactions among quarks, anti-quarks, as well as  
their interactions with gluons.  The ansatz can be translated to the form of the equilibrium distribution functions, 
$ f_{eq}\equiv \lbrace f_{g}, f_{q}, f_{s} \rbrace$ (this notation will be useful later while 
writing the transport equation in both the sector in compact notations)  as follows,
\ba
\label{eq1}
f_{g/q} &=& \frac{z_{g/q}\exp[-\beta E_q]}{\bigg(1\mp z_{g/q}\exp[-\beta E_q]\bigg)},\nn
f_{s} &=& \frac{z_q\exp[-\beta E_q]}{\bigg(1+z_q\exp[-\beta E_q]\bigg)},
\ea
where $E_q=\sqrt{q^2+m^2}$ is the energy of the particle in the thermal bath 
moving with momentum, $q=|\vec{q}|$.  Light quarks and gluons  are taken
as massless $(m=0)$, for strange quark 
$m=m_s (\sim 0.1$ GeV)
and $\beta=T^{-1}$ denotes inverse of the bath temperature. 

We use the notation $\nu_g=2(N_c^2-1)$ for gluonic degrees of freedom,
$\nu_{q}=2\times 2\times N_c\times 2$ for light quarks, $\nu_s=2\times 2 \times N_c \times 1$ 
for the strange quark for $SU(N_c)$. 
Note that we are working 
at zero baryonic chemical potential, therefore quark and antiquark distribution function are the same.
This fact has been taken care of by counting the anti-quark degrees of freedom in $\nu_q$ and $\nu_s$.
Here, we are dealing with $SU(3)$, so  $N_c=3$. Since the model 
is valid in the deconfined phase of QCD (beyond $T_c$), therefore, the mass of the light quarks can be neglected as compared to 
the temperature of the bath.  

Clearly, the determination of $f_g$, $f_q$, and $f_s$ require the temperature 
dependence of the free parameters $z_g$ and $z_q$ respectively. To do this, we rewrite lattice QCD 
pressure symbolically as, \\
\be
P=P_g+P_{qg},
\ee
where $P$ is full (2+1)-flavor lattice QCD pressure, $P_g$ is the contribution from the gluonic action alone
and $P_{qg}$ is the remaining part obtained after subtracting $P_g$ from $P$. The fraction $P_g$, is utilized to define
effective gluonic (quasi-gluons) degrees of freedom, and $P_{qg}\equiv P-P_g$ to effective quark-antiquark degrees of
freedom through an effective Grand canonical partition
function $Z=Z_g\times Z_{qg}$ via the relation $P \beta V=  \ln(Z)$ as follows,
\ba
P_g \beta V&=& \frac{V \nu_g}{8 \pi^3} \int d^3\vec{q}\ \ln \left(1-z_g\exp[-\beta |\vec{q}|]\right)\nonumber\\
P_{qg} \beta V&=& \frac{V}{8 \pi^3}\int d^3\vec{q} \bigg\lbrace \nu_q   \ln \left(1+z_q\exp[-\beta |\vec{q}|] \right)\nonumber\\
&&+\nu_s \ln \left(1+z_q\exp[-\beta\sqrt{|\vec{q}|^2+m^2}]\right)\bigg\rbrace.
\ea
One can read-off $Z_{g}$ and $Z_{qg}$ which are their standard definitions for bosons and fermions.
We know the left hand side of these equation from lattice QCD, and by calculating roots 
of these equations at each temperature, we can determine  $z_g$ and $z_q$
as a function of temperature.  This exercise has been performed numerically. 

It is worth emphasizing that the effective fugacity is not merely a temperature
dependent parameter which encodes the 
hot QCD medium effects. 
Its physical significance is reflected through the modified dispersion relation both in the 
gluonic and quark sector obtained from the thermodynamic relation of energy density and partition function,
$\epsilon=-\partial_\beta \ln(Z)$. One thus find that the effective fugacities modify the single quasi-parton energy
as follows,
\ba
\label{eq2}
\omega_g&=&|\vec{q}|+T^2\partial_T ln(z_g)\nn
\omega_q&=&|\vec{q}|+T^2\partial_T ln(z_q)\nn
\omega_s &=&\sqrt{|\vec{q}|^2+m^2}+T^2\partial_T ln(z_q),
\ea
and this lead to the new energy dispersions for gluons ($\omega_g$), 
light-quark antiquarks ($\omega_q$),
and strange quark-antiquarks,
($\omega_s$). These dispersion relations can be explicated as follows. 
The single quasi-parton energy not only depends upon its momentum
but also gets contribution from the collective excitations of the quasi-partons.
The second term is like the gap in the energy-spectrum due to the presence of 
quasi-particle excitations. This makes the model more in the spirit of the Landau's theory of Fermi -liquids.
For a  detailed discussions of the interpretation and physical significance of $z_g$, and $z_q$, we refer the reader 
to~\cite{chandra_quasi}. There are other quasi-particle descriptions in the literature, those could be 
characterized as, effective mass models ~\cite{effmass1,effmass2},
effective mass models with gluon condensate~\cite{effmass_glu}, and 
effective models with Polyakov loop~\cite{effmass_pol}. 
Our model is fundamentally distinct from all these models.
Another crucial point is regarding the definition of the 
energy momentum tensor, $T^{\mu\nu}$. As described in~\cite{dusling}, in the presence of non-trivial temperature 
dependent energy dispersion (as in all these quasi-particle models), we need to modify the definition of the 
$T^{\mu\nu}$ so that the trace anomaly effects in QCD can be accommodated in the definition. The 
modified $T^{\mu\nu}$ for the effective mass models is obtained in~\cite{dusling}, and for the current model 
in ~\cite{chandra_new}.

\subsection{Viscous QGP: modification to thermal distributions}
Viscosities usually modifies the equilibrium thermal distributions as follows,
\be
f =f_{eq}+{\delta f}_{\eta} +{\delta f}_{\zeta}.
\ee
where $f$ jointly denote the viscous modified thermal distribution 
for  gluons, quark and anti-quarks (in the present case
gluons, light quarks and strange quark), therefore, $f\equiv \lbrace f_{gg}, f_{qq}, f_{ss}\rbrace$.
For the shear and bulk viscous corrections, ${\delta f}_{\eta}$ and ${\delta f}_{\zeta}$ respectively, 
we utilize the results of~\cite{dusling} by using only first order 
terms in the expansion of shear and bulk part of the stress tensor, and choosing the 
local rest frame of the fluid.  Precisely, we have minimally extended the 
expressions in the present case by substituting the thermal distributions of quarks and 
gluons as the form obtained in the effective fugacity quasi-particle model.
This is solely based on the fact that the quasi-partons in the present case are non-interacting up to their 
respective effective fugacities that encode all the interactions. This leads to the following expressions,
  \ba
f_{gg}&=&f_g+ \frac{f_g [1+f_g]}{2T^3\tau} \bigg[(\frac{|\vec{q}|^2}{3}-q_z^2)\frac{\eta}{s}
 +\frac{2 |\vec{q}|^2}{5} \frac{\zeta}{s}\bigg]\\
\label{zg1}
 f_{qq}&=&f_q+ \frac{f_q [1-f_q]}{2T^3\tau} \bigg[(\frac{|\vec{q}|^2}{3}-q_z^2)\frac{\eta}{s}
 +\frac{2 |\vec{q}|^2}{5} \frac{\zeta}{s}\bigg]\\
\label{zg2}
 f_{ss}&=&f_s+ \frac{f_s [1-f_s]}{2T^3\tau} \bigg[(\frac{|\vec{q}|^2}{3}-q_z^2)\frac{\eta}{s}
 +\frac{2 |\vec{q}^2|}{5} \frac{\zeta}{s}\bigg],
\label{zg}
 \ea
where $\eta$ and $\zeta$ are the shear and bulk viscosities of the QGP, $s$ is the entropy density
and  $\tau$ is the thermalization time of the QGP. 
We will employ the same temperature independent 
values of $\eta/s$ and $\zeta/s$ while comparing the effects of 
ideal and non-ideal EoS with viscosities. 
For simplicity, same constant values are also taken for $\eta/s$ and $\zeta/s$ for gluons, and quarks.

There are several interesting reports on the 
estimation of a small value of $\eta/s$ for QGP, and the 
emergence of near perfect fluid picture~\cite{shrvis,bmuller,chandra_eta1,chandra_eta2}.
It has been realized in the recent past that bulk viscosity of the QGP is significant  
near the transition temperature,  $T_c$~\cite{chandra_new,khaz,buch,chandra_bulk}. 
This is attributed to the large value of the interaction measure 
near $T_c$~\cite{khaz}.  Before we present our results on the effects of $\eta$ and $\zeta$
on the phase space distribution it is imperative to mention viscous correction to phase space 
distribution is based on momentum expansion, which will be valid when the momentum (q) 
is small~\cite{DTeaney}. It should be noted here that the momentum of the thermal particles are
integrated out in evaluating the drag and diffusion coefficients of HQ and the thermal phase
space factor become small at $q\sim 3-4$ GeV for the temperature range considered here.
In the present work the drag of the HQ are evaluated as a function of its momentum and 
temperature of the bath.

The mass of the strange quarks does not play 
any prominent role at high temperature ($T\geq T_c$) because $m/T<1$. In fact, the mass effects are 
negligibly small (of the order of $(m/T)^2$). Therefore, 
the strange quark mass can be ignored which makes the numerical calculations simpler.

Let us set the notation here to distinguish the results obtained for  
the cases with ideal EoS and LEoS as well as  for zero and non-zero values of $\eta$ and $\zeta$.
The quantity $\gamma_{Id}$ stands for the drag coefficient for ideal case,  $\gamma^{eos}$
denotes that for the LEoS, and $\gamma_{\eta}$ $(\gamma_\zeta)$ stands for drag coefficients when the effect of $\eta$
 $(\zeta)$ is included through the phase space of the bath particles. 
$\gamma_{\eta+\zeta}$ denotes the drag coefficient while both $\eta$ and $\zeta$
are taken in to account with ideal QGP EoS and as ${\gamma}^{eos}_{\eta+\zeta}$ in the case of the realistic QGP EoS.
Similarly, the quantity $D^{eos}_{\eta+\zeta}$ denotes the  HQ diffusion coefficient while LEoS and both $\eta$ and $\zeta$
are taken in to account, the notation $D_{Id}$ denotes same for ideal EoS without viscous effects.

\begin{figure}[ht]
\vspace{8mm}
\begin{center}
\includegraphics[scale=0.35]{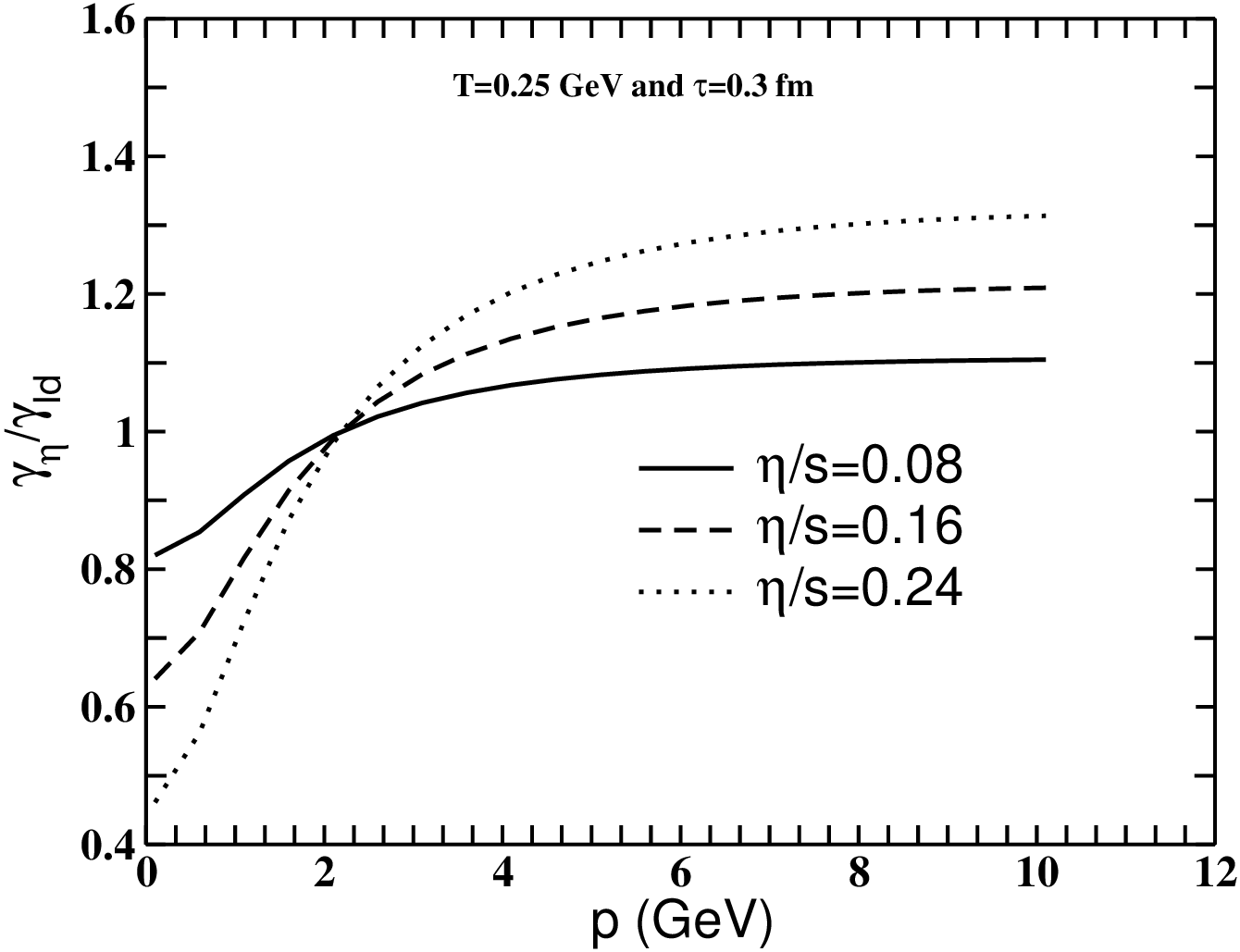}\vspace{10mm}
\includegraphics[scale=0.35]{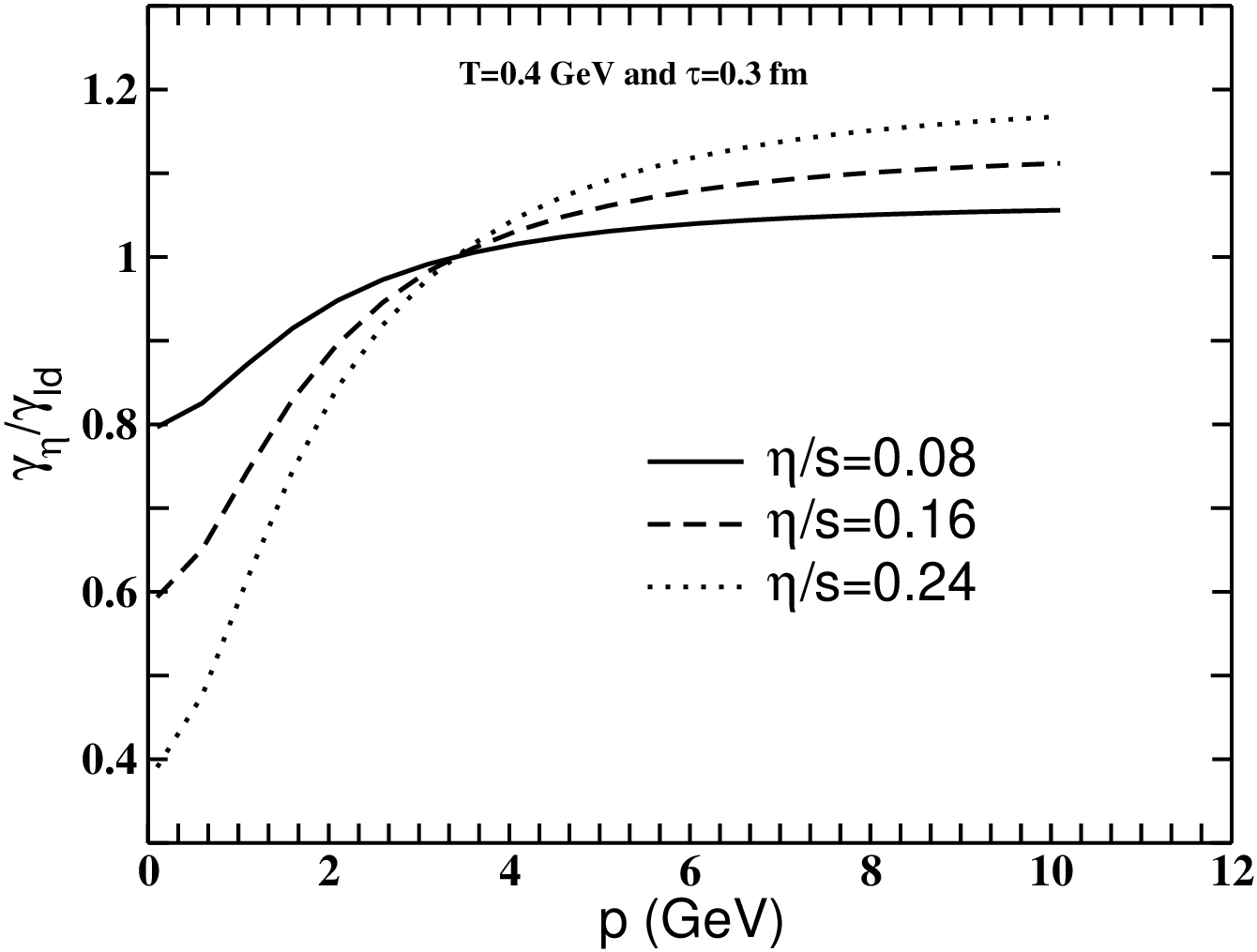}
\caption{ Variation of the ratio $\gamma_{\eta}/\gamma_{Id}$
as a function of momentum with different values of $\eta/s$ at a given 
temperature are shown (upper panel). Same quantity at a different 
temperature is depicted in the lower panel}
\label{fig1}
\end{center}
\end{figure}
\section{Results and discussions}
\begin{figure}[ht]
\vspace{8mm}
\begin{center}
\includegraphics[scale=0.35]{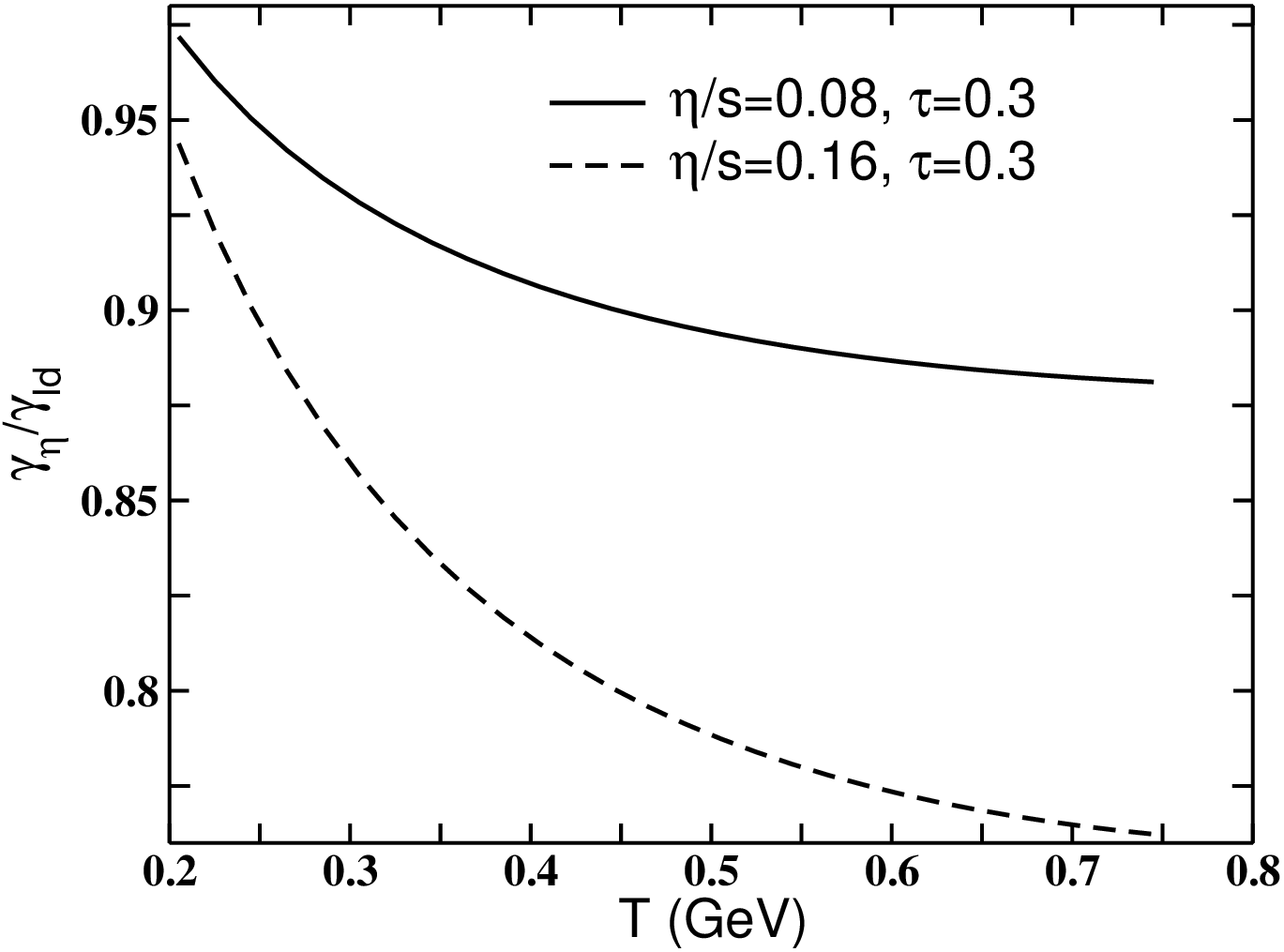}\vspace{10mm}
\includegraphics[scale=0.35]{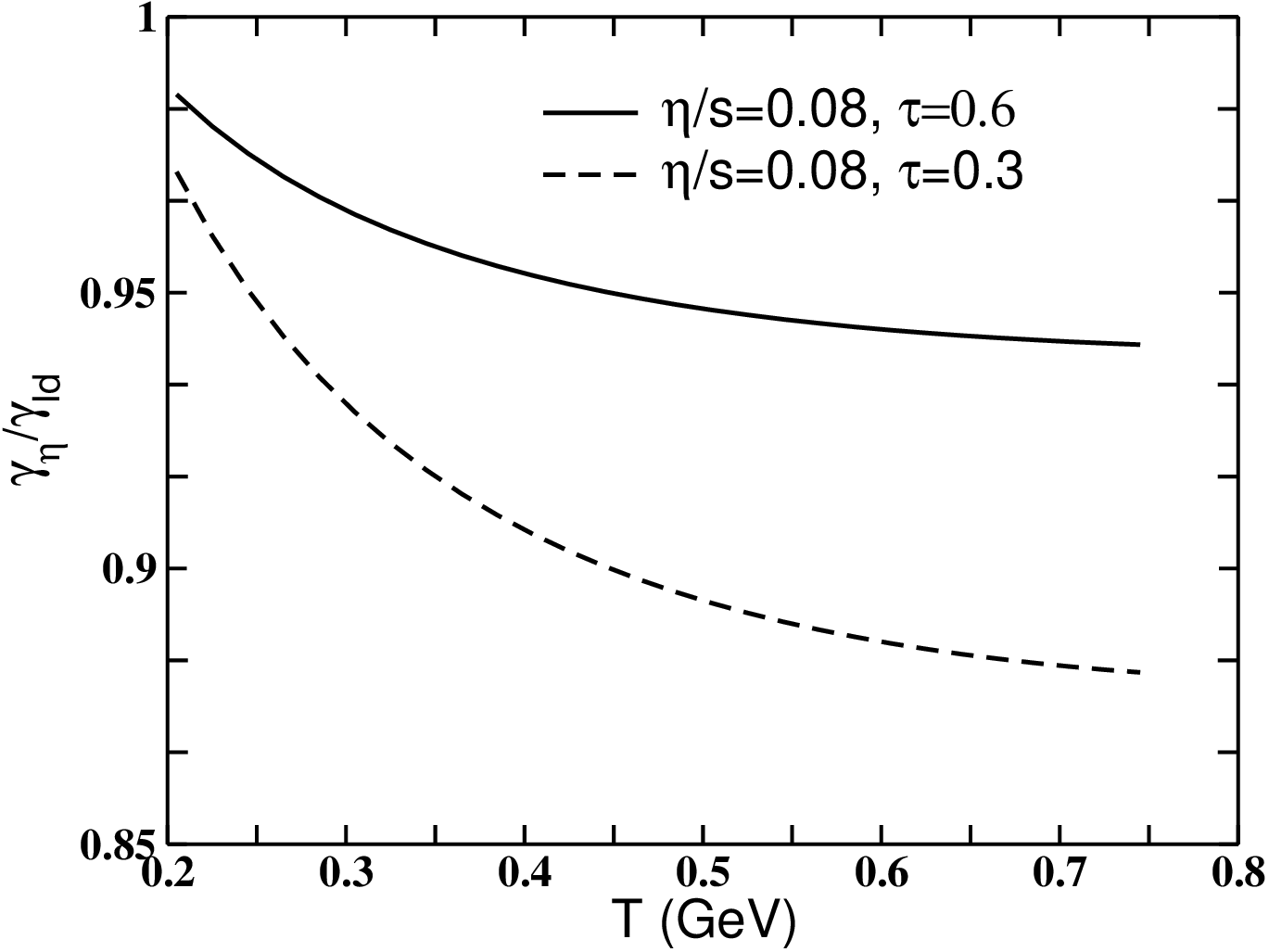}
\caption{The variation of the ratio $\gamma_{\eta}/\gamma_{Id}$
as a function of temperature with different values of $\eta/s$ at a given 
thermalization time, $\tau$ are shown (upper panel). Same quantity with the variation of 
$\tau$ for fixed $\eta/s$ is shown in the lower panel}
\label{fig2}
\end{center}
\end{figure}

\begin{figure}[ht]
\vspace{8mm}
\begin{center}
\includegraphics[scale=0.35]{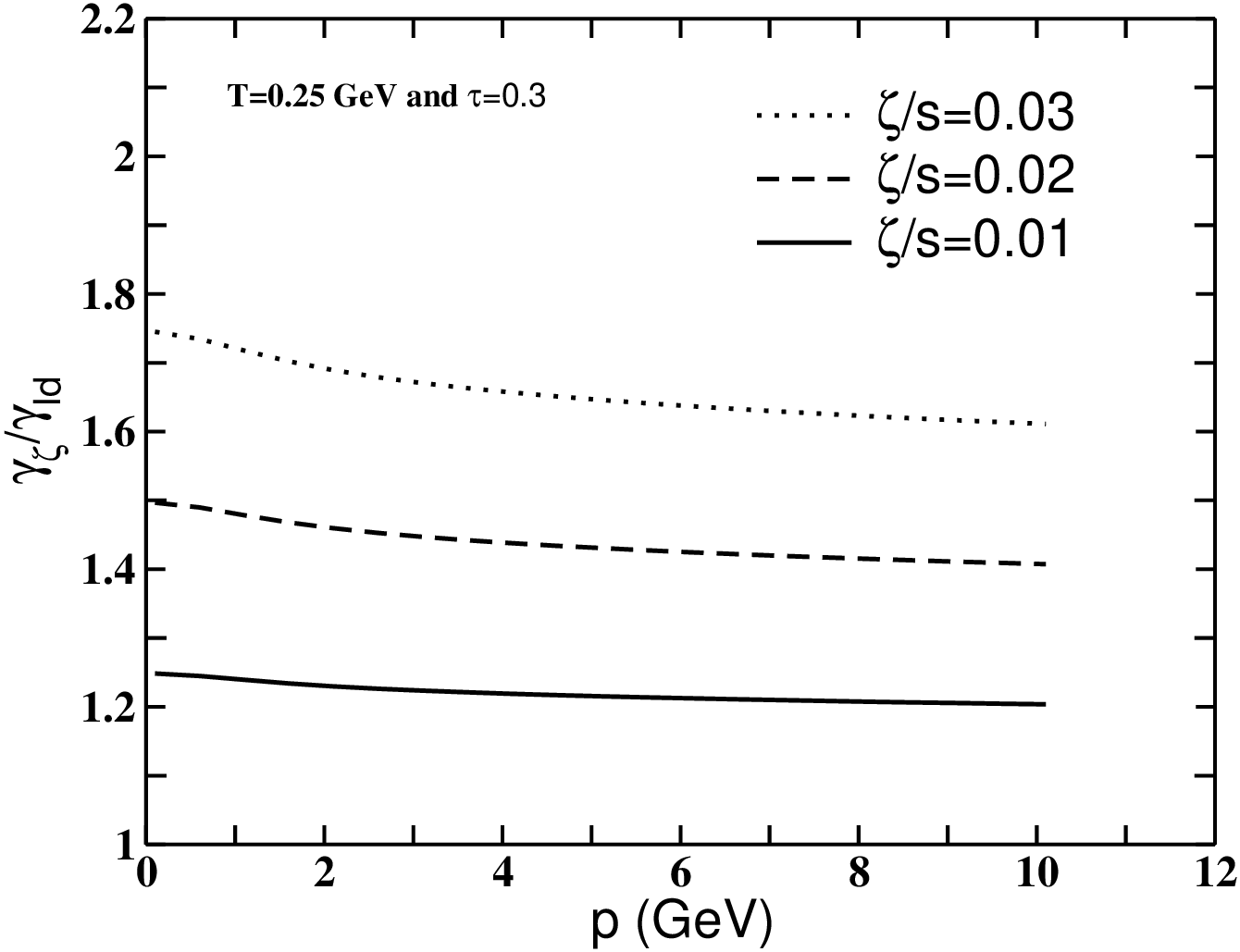}\vspace{10mm}
\includegraphics[scale=0.35]{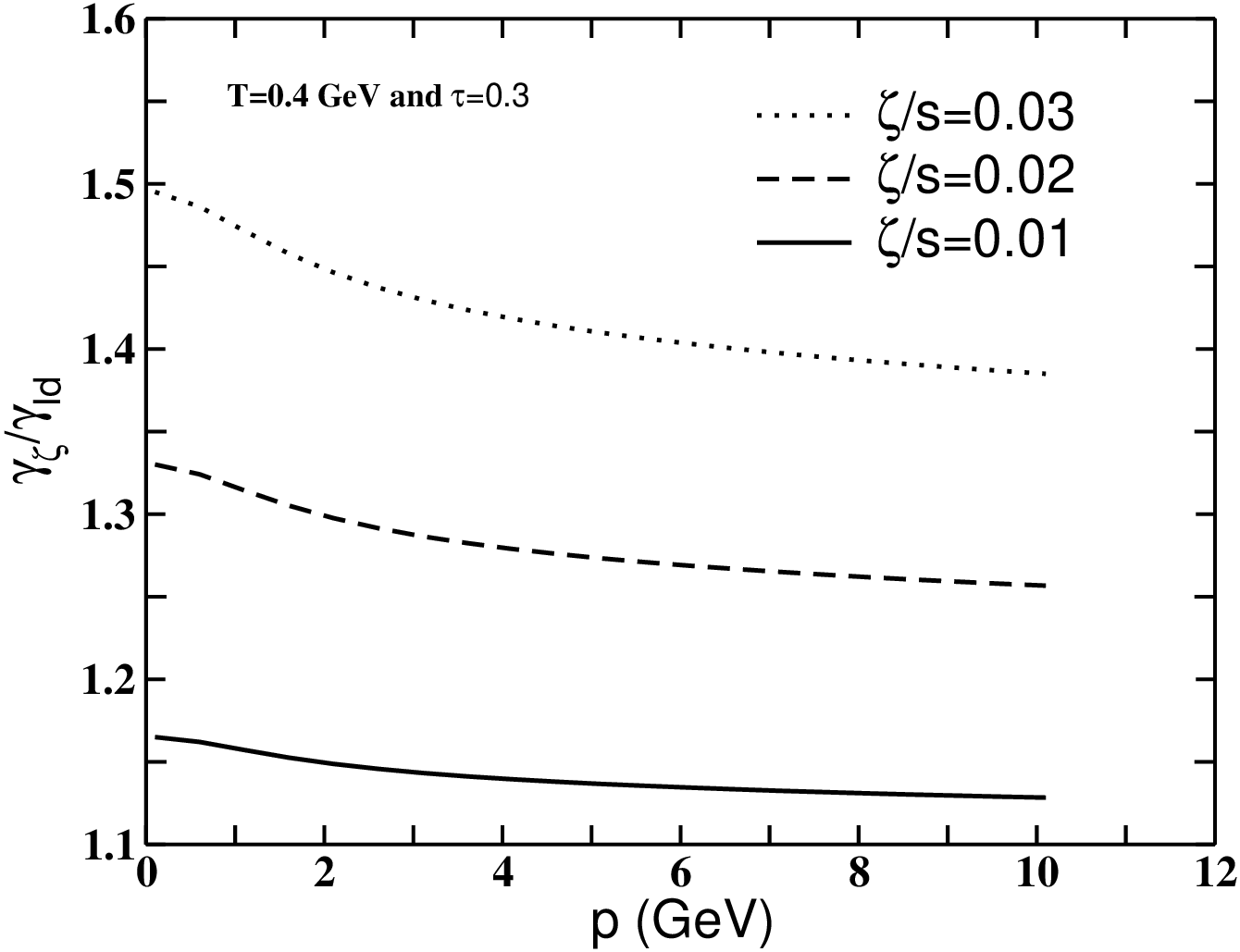}
\caption{The variation of the ratio $\gamma_{\zeta}/\gamma_{Id}$
as a function of momentum with different values of $\zeta/s$ at a given 
temperature are shown (upper panel). Same quantity at a different 
temperature is shown in the lower panel}
\label{fig3}
\end{center}
\end{figure}

\begin{figure}[ht]
\vspace{8mm}
\begin{center}
\includegraphics[scale=0.35]{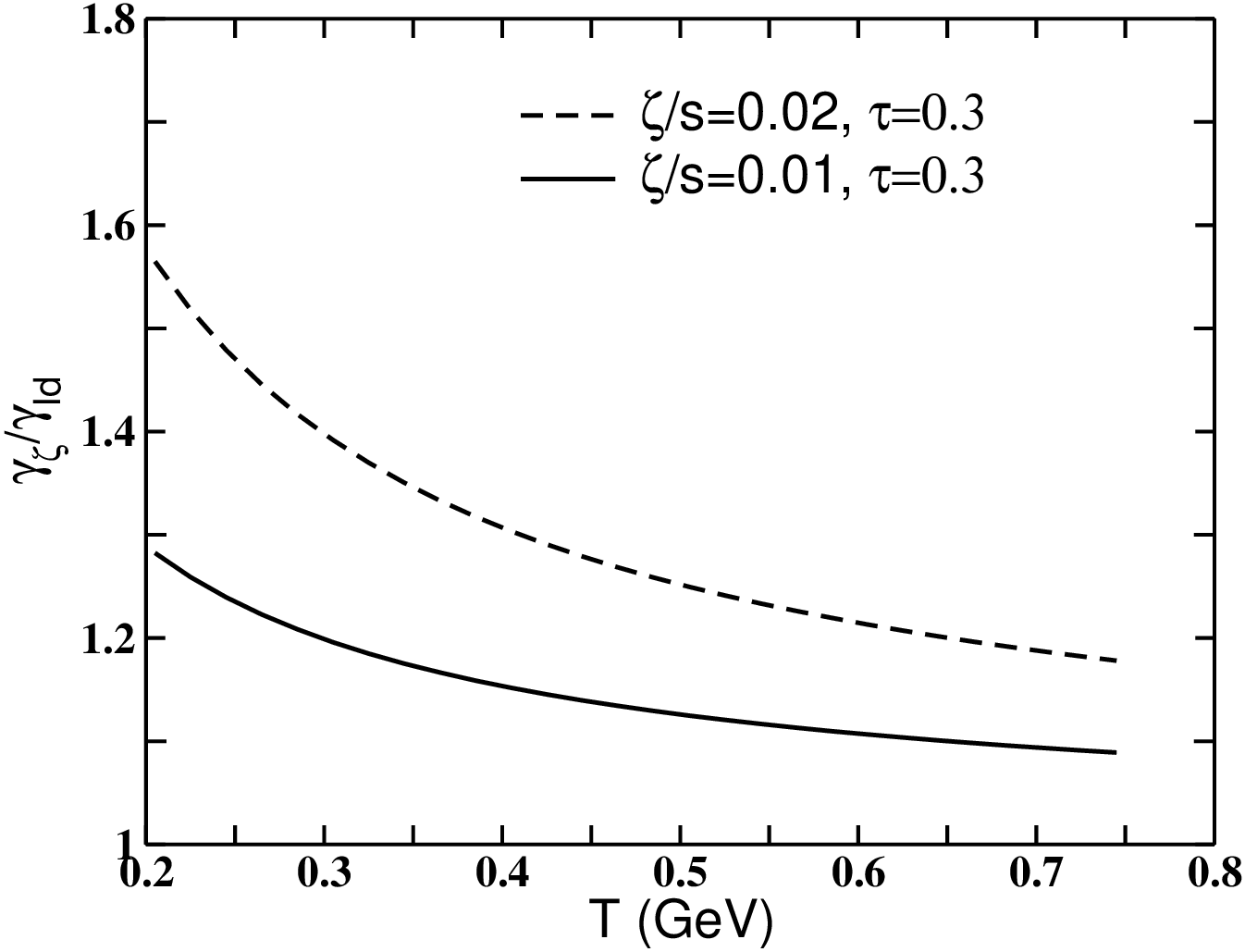}\vspace{10mm}
\includegraphics[scale=0.35]{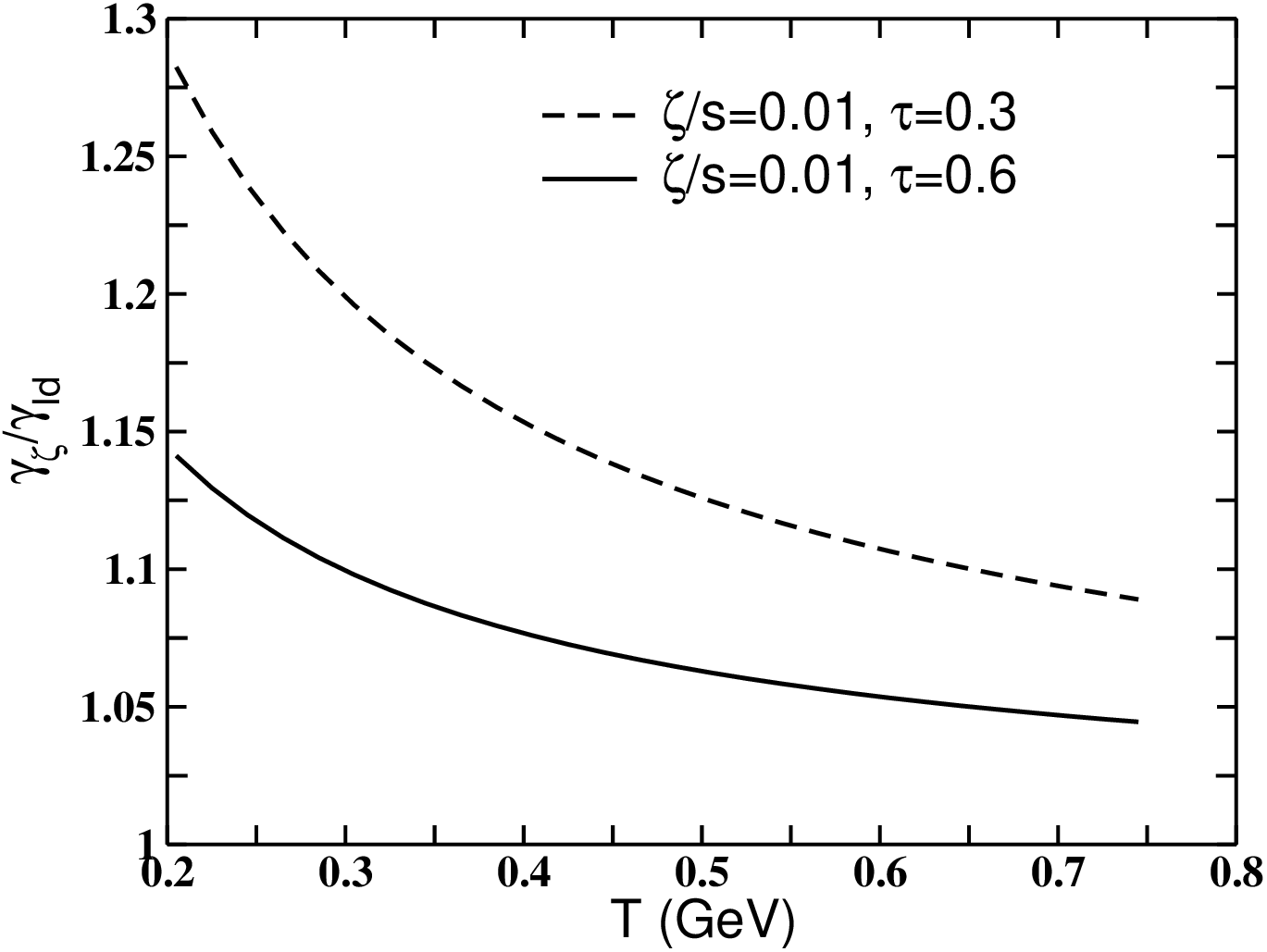}
\caption{Upper panel: The variation of the ratio $\gamma_{\zeta}/\gamma_{Id}$ 
as a function of temperature with different values of $\zeta/s$ at a given 
value of thermalization time $\tau$ are depicted. Lower panel: same quantity has been plotted for
fixed $\zeta$ but with varying $\tau$.}
\label{fig4}
\end{center}
\end{figure}

\begin{figure}[ht!]
\vspace{8mm}
\begin{center}
\includegraphics[scale=0.35]{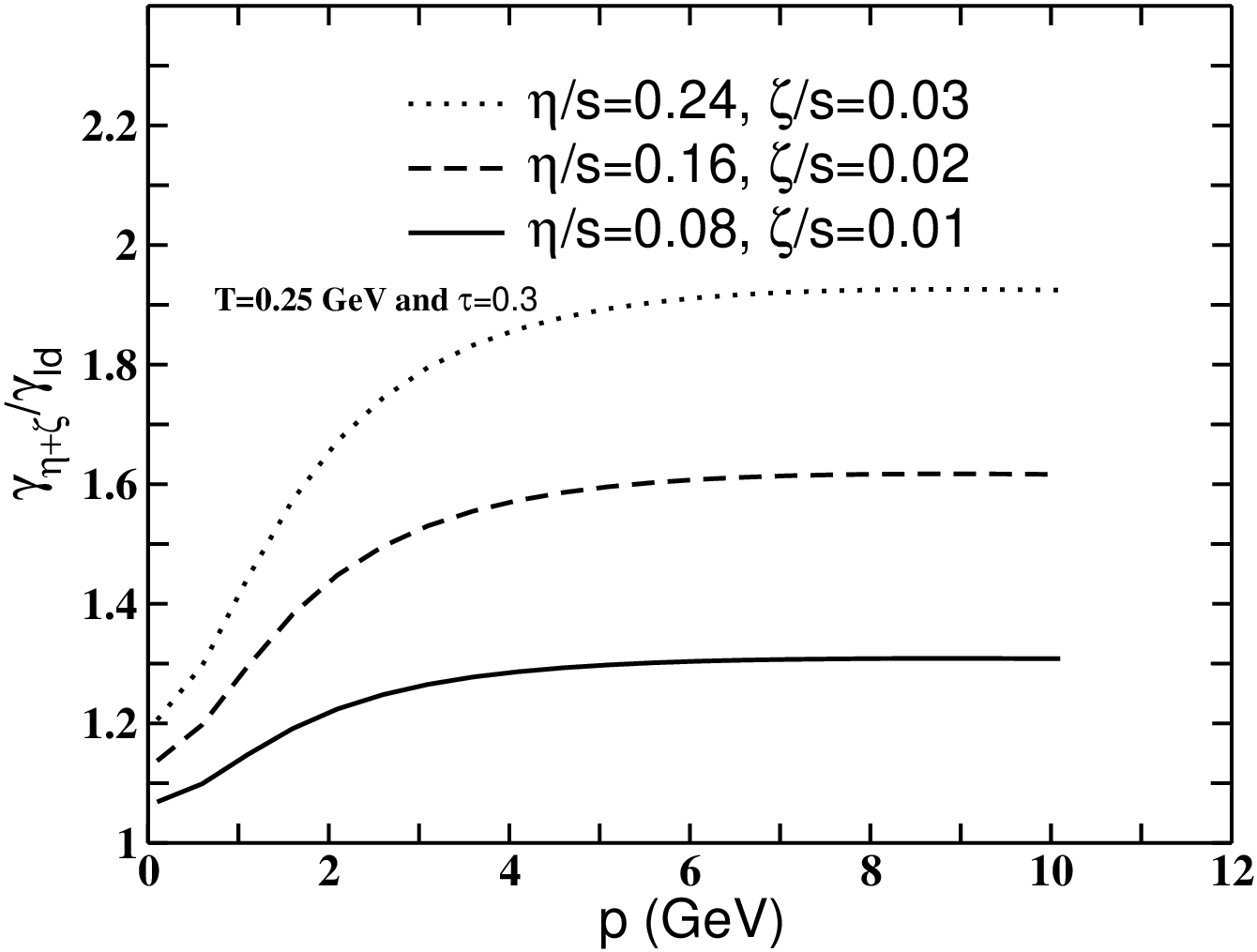}\vspace{10mm}
\includegraphics[scale=0.35]{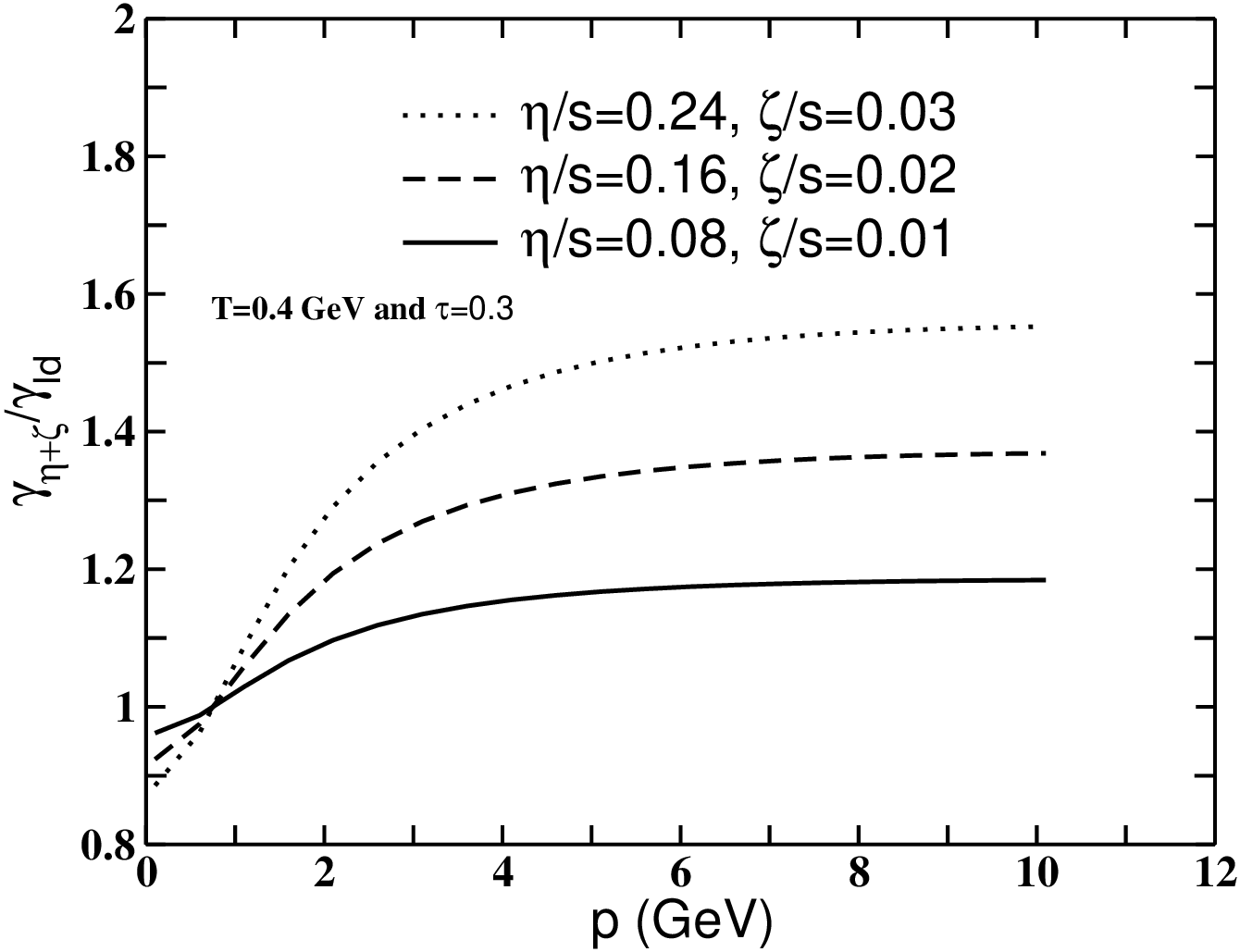}
\caption{ The variation of the ratio $\gamma_{\eta+\zeta}/\gamma_{Id}$
as a function of momentum with different values of $\eta/s$ and $\zeta/s$ at a given 
temperature are shown (upper panel). Same quantity is plotted at a different 
temperature in the lower panel.}
\label{fig5}
\end{center}
\end{figure}

\begin{figure}[ht!]
\vspace{8mm}
\begin{center}
\includegraphics[scale=0.35]{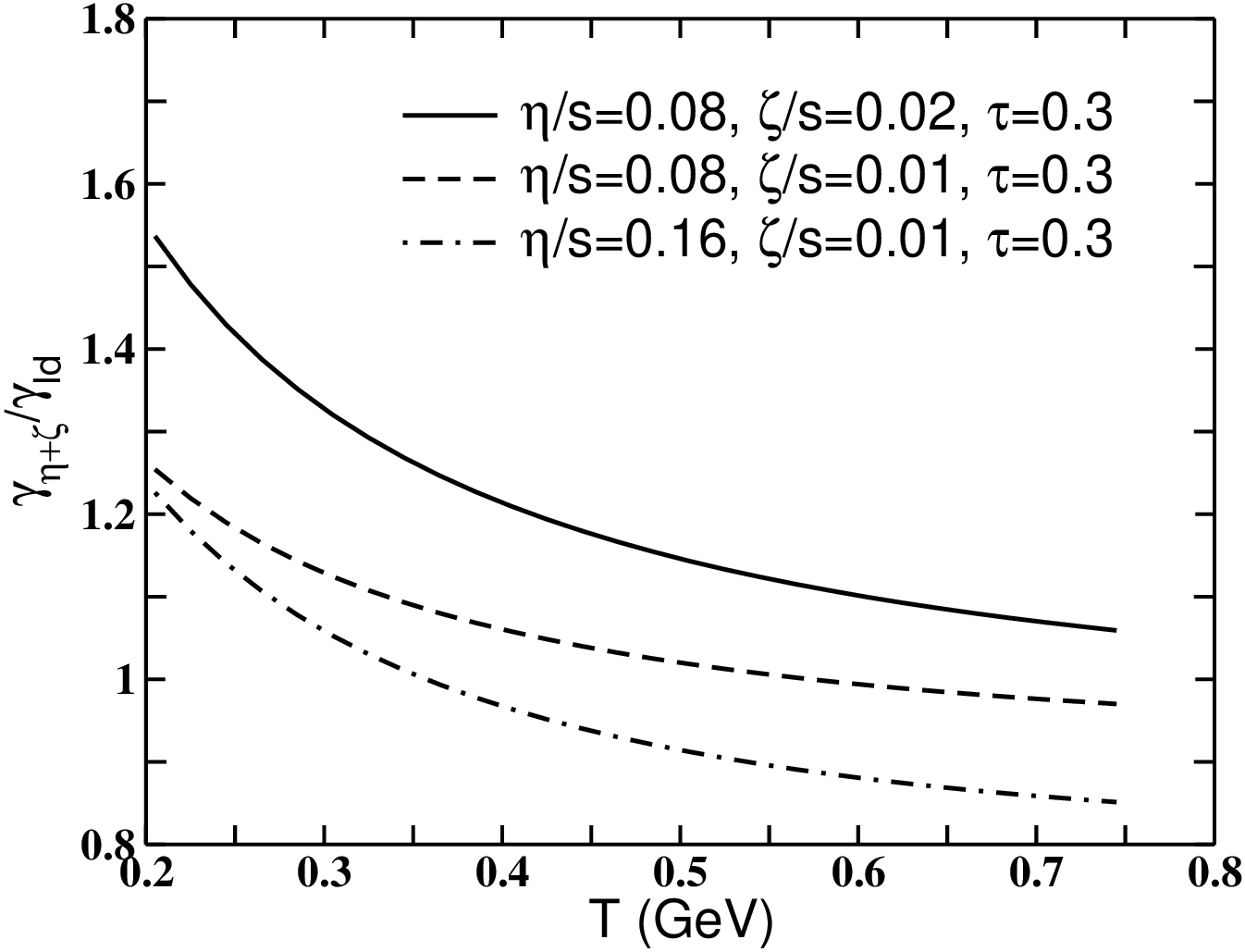}
\caption{$\gamma_{\eta+\zeta}$ stands for drag coefficients when the effects of
both the shear and bulk viscosities are included  through the thermal distribution
of the bath particles in evaluating the drag 
coefficient. The variation of the ratio of the drag coefficients 
as a function of temperature with different values of $\zeta/s$,$\eta/s$ and $\tau$ are shown.}
\label{fig6}
\end{center}
\end{figure}

\begin{figure}[ht!]
\begin{center}
\vspace{8mm}
\includegraphics[scale=0.35]{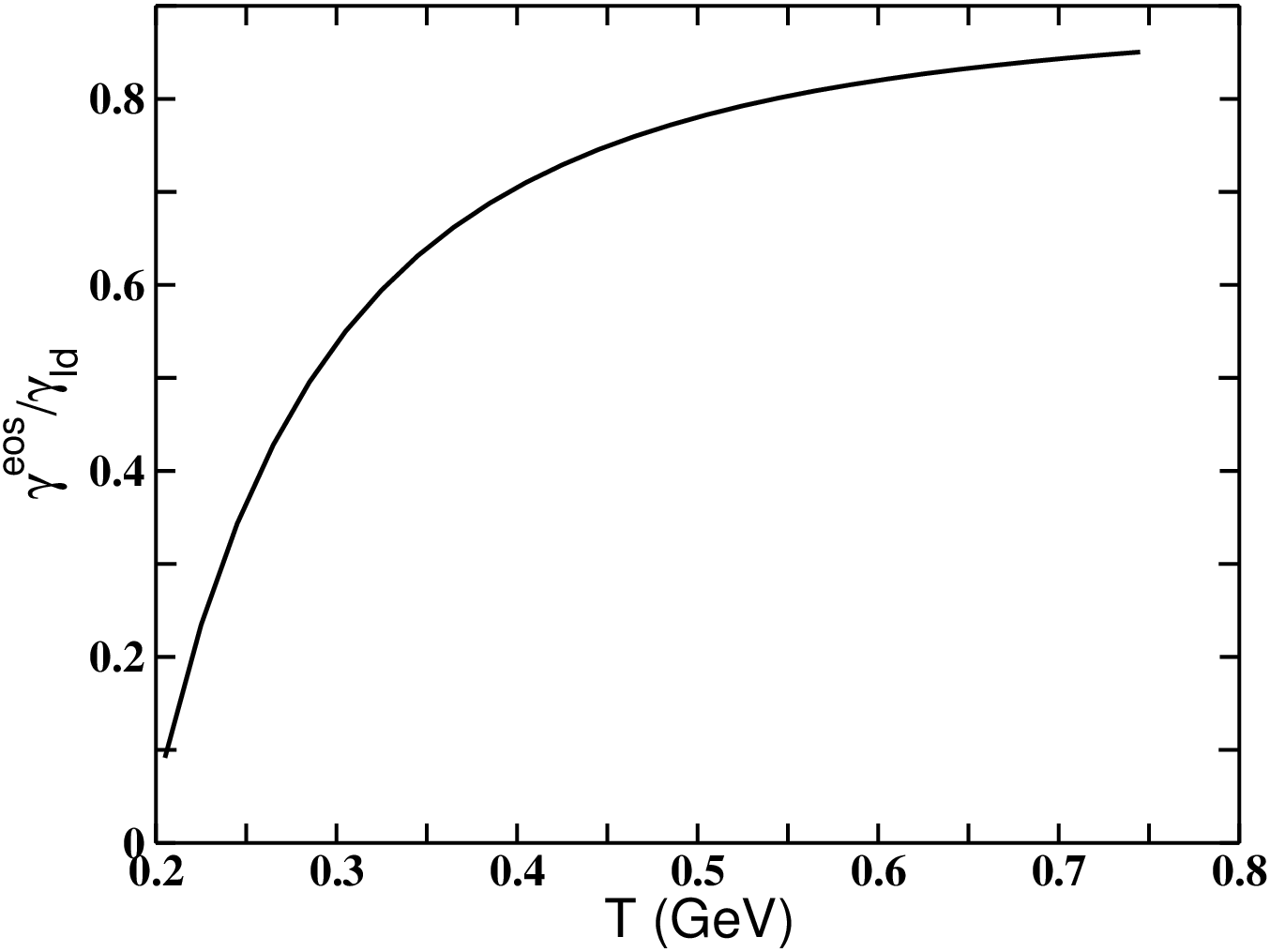}
\caption{$\gamma^{eos}$ stands for drag coefficients when effect of EoS 
is considered in evaluating the drag coefficient. 
The variation of the ratio of the drag coefficients 
as a function of temperature are shown.}
\label{fig7}
\end{center}
\end{figure}

\begin{figure}[ht!]
\vspace{8mm}
\begin{center}
\includegraphics[scale=0.35]{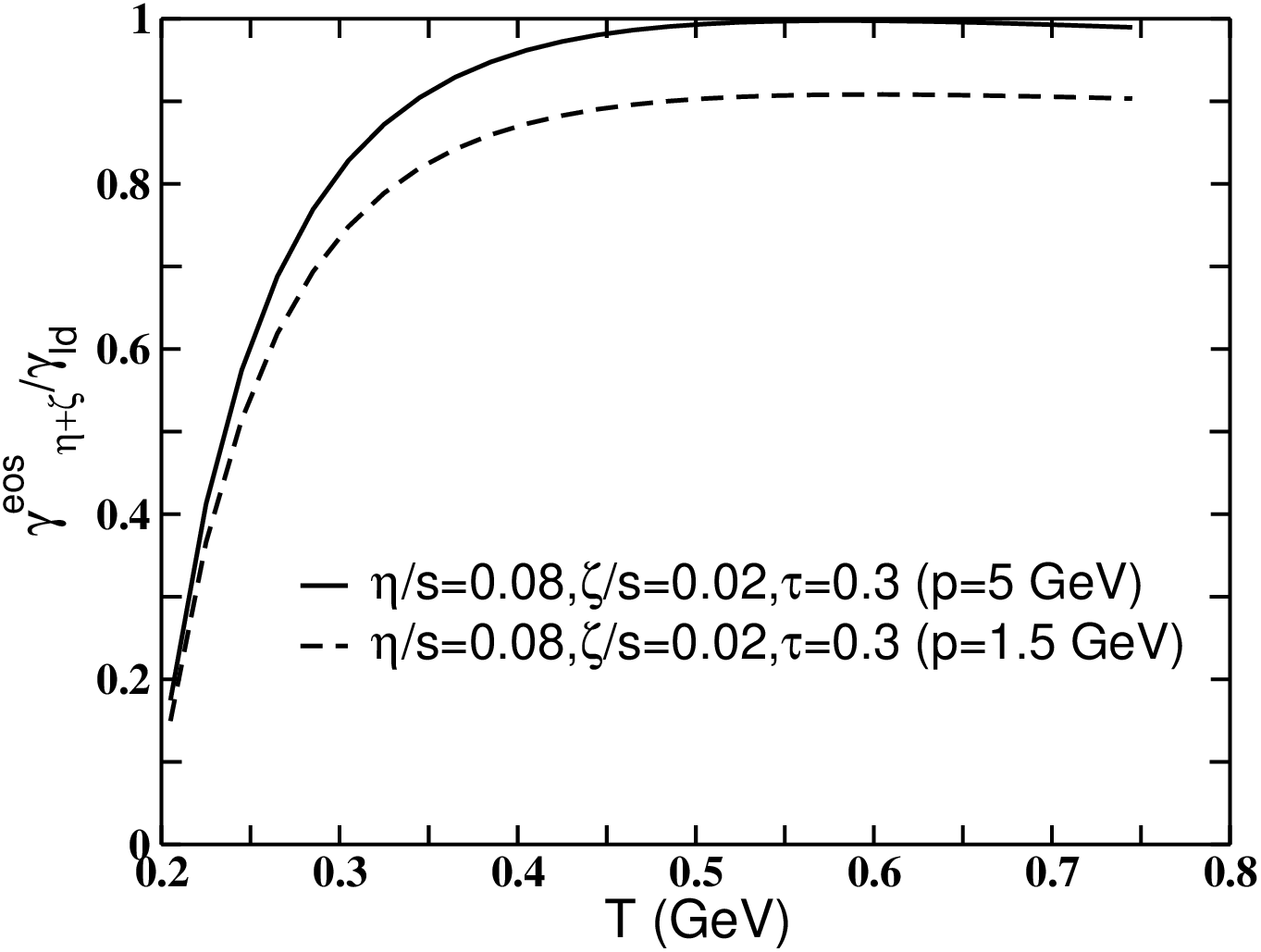}
\caption{The variation of the quantity $\gamma^{eos}_{\eta+\zeta}/\gamma_{Id}$ as a function of temperature.}
\label{fig8}
\end{center}
\end{figure}

\begin{figure}[ht!]
\vspace{8mm}
\begin{center}
\includegraphics[scale=0.35]{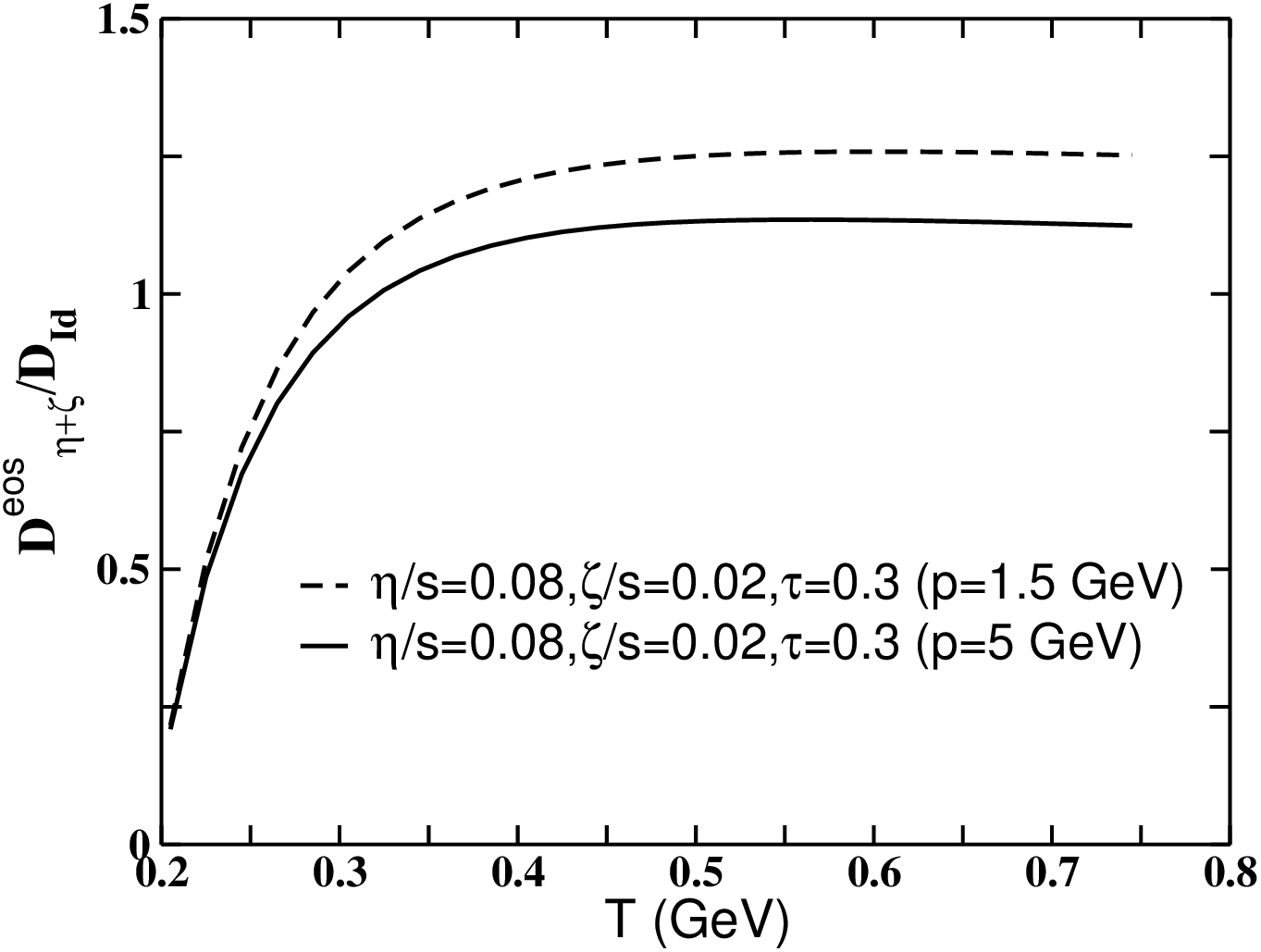}
\caption{
The variation of the quantity $D^{eos}_{\eta+\zeta}/D_{Id}$ as a function of temperature.}
\label{fig9}
\end{center}
\end{figure}

In this section, we display the momentum and temperature dependence of the HQ
drag coefficient employing the realistic EoS for the QGP
and the dissipative environment mainly induced by the
shear and bulk viscosities. Our focus is mainly on the
impact of the effects of the viscosity and the realistic
EoS on the HQ transport coefficients.
Our aim is served by the ratio of drag coefficients evaluated with 
ideal (excluding dissipation) and LEoS (including 
viscous effects). These effects are demonstrated through the 
results displayed in the Figs. \ref{fig1}-\ref{fig9}.

In Fig~\ref{fig1} (upper panel), the effect of shear viscosity on drag coefficient has been depicted as a 
function of momentum at $T=0.25$ GeV through the ratio of the drag coefficients 
with and without shear viscous effects with ideal EoS. The drag coefficient for zero shear viscosity  
is higher compared to the case of nonzero shear viscosity in the low momentum range. 
A reduction in the ratio of the drag coefficients for low $p$ is obtained because 
of the decrease in the phase space factor of the particles in the thermal 
bath due to the possible negative contribution from the shear viscous corrections 
as indicated in Eqs.~\ref{zg1}-\ref{zg}. 
The situation get reversed at higher momenta ($p> 2$ GeV).  
In Fig~\ref{fig1} (lower panel) the same quantity 
has been plotted at a higher temperature ($T=0.4$ GeV). 
The nature of the variation of the ratio remains similar. 
However, the magnitude of the ratio reduces with increasing temperature, indicating
the modulation of relative drag coefficient induced by $\eta/s$.

In Fig~\ref{fig2} (upper panel), the effect of shear viscosity on drag
coefficient has been depicted as a function of temperature $(>T_c)$
 for momentum, $p=1.5$ GeV.
The drag coefficient for zero shear viscosity is
higher compared to the case of nonzero shear viscosity
in the entire temperature range considered here. 
This is because of the fact that  we have performed the calculation at low momentum.
The same quantity is displayed in Fig~\ref{fig2} (lower panel), 
as a function of temperature for  different values of thermalization time at a given value 
of $\eta/s$. It is observed that effects of shear viscosity get milder with the increase in  
thermalization time (as expected from the terms appearing as the coefficients of $\eta/s$).

In Fig~\ref{fig3} (upper panel), the effect of the 
bulk viscosity on drag coefficient (relative to the ideal case) has been displayed as a 
function of momentum for $T=0.25$ GeV. It is observed that the  drag coefficient get enhanced 
in the presence of bulk viscosity. 
This enhancement in the ratio of the drag coefficients for low $p$ may arise 
from the increase in the phase space factor of the particles in the thermal 
bath due to the positive bulk viscous corrections as 
indicated in Eqs.~\ref{zg1}-\ref{zg}. 
That is, since the  term appearing as the coefficient of $\zeta/s$ is always positive definite, the
presence of bulk viscosity will enhance the  phase space density and hence the drag coefficients.
If we increase the $\zeta/s$, drag coefficient is also enhanced. 
In Fig~\ref{fig3} (lower panel), the same quantity has been plotted for $T=0.4$ GeV. 
It is observed that the ratio reduces with increase in temperature.

In Fig~\ref{fig4} (upper panel), the effect of the 
bulk viscosity on drag coefficient (relative to the ideal case) has been displayed as a 
function of temperature for $p=1.5$ GeV. 
The effects of thermalization time is displayed in Fig~\ref{fig4} (lower panel).
It is found that the effect of bulk viscosity 
becomes weaker at larger thermalization time. 

Considering both the shear and 
the bulk viscosities, we compute the drag coefficient ($\gamma_{\eta+\zeta}$) and 
study its variation with the momentum for temperatures $T=0.25,\, 0.40 $ GeV.
The results are depicted in the upper and lower panels of Fig.~\ref{fig5}.
We observe that the ratio $\gamma_{\eta+\zeta}/\gamma_{Id}$ 
get enhanced with increase in momentum. 
The increase is faster at low momentum, slower at higher 
momentum.  The magnitude of the quantity $\gamma_{\eta+\zeta}/\gamma_{Id}$ is lower at higher 
value of $T$ ($0.4$ GeV) through out the range of the momentum considered here.
Interestingly, the low momentum  domain is dominated by shear
viscous effects, whereas high momentum part is dominated by bulk viscous effects. 

In Fig~\ref{fig6}, the temperature variation (for $p=1.5$ GeV) of the drag coefficient
relative to the ideal case has been shown in the presence of both $\eta/s$ and $\zeta/s$. 
It is the interplay of these two dissipative effects that dictates the behavior of the ratio 
as a function of $T$.

The ratio of the drag coefficients is plotted against $T$ in Fig~\ref{fig7}  to 
understand the effects of EoS.  The effects of interactions 
contained in LEoS suppresses the phase space distributions of the bath particle
which is seen through the  reduction of the ratio of drag  
coefficients at low temperature in Fig.~\ref{fig7}. The ratio increases 
with $T$ because the effects of interaction at high $T$ becomes weaker. 
Ideally the ratio will become unity  in the Stefan-Boltzmann limit.  

In Fig~\ref{fig8} combined effects of both shear and bulk viscosities along with the EoS have been presented as a function 
of temperature at given values of $\tau$, $\eta/s$, $\zeta/s$ at two different momentum. The reduction in the drag 
is significant at the temperature domain relevant of nuclear collisions at RHIC and LHC energies for the 
low momentum case. Therefore, this results will have crucial effects on the nuclear suppression of heavy flavors 
measured at these energies. 

The effects of phase space corrections due to non-zero shear and bulk 
viscosities and realistic EoS on diffusion coefficient ($D\equiv B_0$) have been
shown explicitly in Fig.~\ref{fig9}. It is observed that the 
ratio of diffusion coefficients, $D^{eos}_{\eta+\zeta}/D_{Id}$ is significantly 
smaller than unity for low $T (<300$ MeV). The effects of realistic EoS, {\it i.e.} the
LEoS in the present case, is found to play the most dominant role at low $T$. At higher
$T$ the effects of LEoS becomes weaker and the effects viscous corrections to phase
space factors start playing significant role.

\section{SUMMARY AND CONCLUSIONS}
The drag and diffusion coefficients of the HQs propagating through a viscous system of
quarks and gluons have been evaluated by incorporating the corrections to the phase space
factors arising from the non-zero viscosities of the system which restrain the system slightly
away from the equilibrium. This corrections seem to be significant for the temperature range
expected to be attained in the system produced at HIC at RHIC and LHC collision energies. 
The effect of realistic EoS on the drag has also been ascertained. The combined effects of viscosities
and the EoS reduces the drag significantly at moderate momentums. 
These results will have crucial consequences on the heavy
ion phenomenology at RHIC and LHC energies, which is the matter of future investigations.


\vspace{2mm}
\section*{Acknowledgements} 
SKD thanks Payal Mohanty for useful discussions and acknowledges FAIR center, VECC Kolkata, India 
for financial support during his visit to VECC. VC sincerely acknowledges the financial support from INFN, Italy
for awarding an INFN postdoctoral fellowship at its Firenze section. SKD acknowledges the support by the ERC StG 
under the QGPDyn Grant no. 259684.


\section*{Appendix}
In this appendix, we quote the invariant amplitude for the elastic processes
used for evaluating the drag and diffusion coefficients of charm quark 
propagating through  QGP. 

The invariant amplitude,
$\bigg|\mathcal{M}\bigg|^2_{gc\rightarrow gc}$
 for the process $gc\rightarrow gc$ is given by:

\begin{eqnarray}
\bigg|\mathcal{M}\bigg|^2_{(gc\rightarrow gc)}={\pi^2 \alpha_s^2} 
\bigg\lbrack \frac{32(s-M^2)(M^2-u)}{t^2} 
+\frac{64}{9×}\frac{(s-M^2)(M^2-u)+2M^2(s+M^2)}{(s-M^2)^2×} \nonumber  \\ 
+\frac{64}{9×}\frac{(s-M^2)(M^2-u)+2M^2(M^2+u)}{(M^2-u)×}+
\frac{16}{9×}\frac{M^2(4M^2-t)}{(s-M^2)(M^2-4)×} \nonumber  \\   
+16\frac{(s-M^2)(M^2-u)+M^2(s-u)}{t(s-M^2)×}-16\frac{(s-M^2)(M^2-u)-M^2(s-u)}{t(M^2-u)×}\bigg\rbrack 
\end{eqnarray}
Similarly for the process, $qc\rightarrow qc$, the square of the invariant matrix element,   
$\bigg|\mathcal{M}\bigg|^2_{qc\rightarrow qc}$ is given by:

\begin{eqnarray}
\bigg|\mathcal{M}\bigg|^2_{(qc\rightarrow qc)}\,=\,\frac{64 \pi^2 
\alpha_s^2}{9}
\left[\frac{\left(M^2-u\right)^2+\left(s-M^2\right)^2+2M^2t}
{t^2}\right].
\end{eqnarray}

\section{References}
  
\end{document}